\pdfoutput=1
\documentclass[journal]{IEEEtran}
\usepackage[ruled,vlined,linesnumbered]{algorithm2e}
\usepackage{amsmath}
\usepackage{amssymb}
\usepackage{mathtools}
\usepackage{amsthm}
\usepackage{color}
\usepackage[dvipsnames]{xcolor}
\usepackage{hyperref}
\usepackage{multirow}
\usepackage{array}
\usepackage{diagbox}
\usepackage{colortbl}
\usepackage{soul}
\usepackage{calc}
\usepackage{ulem}
\usepackage{cite}
\usepackage{hyperref}
\newlength{\DepthReference}
\newlength{\HeightReference}
\newlength{\Width}%
\newcommand{\GreyBox}[2][lightgray]%
{%
    \settowidth{\Width}{#2}%
    \setlength{\fboxsep}{0pt}%
    \colorbox{#1}%
    {%      
        \raisebox{-\DepthReference+0.5pt}%
        {%
           \parbox[b][\HeightReference+\DepthReference][c]{\Width}{\centering#2}%
        }%
    }%
}

\theoremstyle{remark}

\theoremstyle{definition}

\newcommand{\tred}[1]{\textcolor{red}{#1}}
\definecolor{weakgray}{gray}{0.9}

\definecolor{darkolive}{rgb}{0.0, 0.5, 0.0} %dark olivegreen
\newcommand{\tgreen}[1]{\textcolor{darkolive}{#1}}

\newcolumntype{?}{!{\vrule width 1pt}}

\begin{document}

\title{FDD Massive MIMO - Antenna Duplex Pattern an-Reciprocity : A Missing Brick}
\author{\IEEEauthorblockN{Patrick C.F. Eggers, \textit{Member}, IEEE, and Stanislav S. Zhekov } \\
\date{November 2020}
\thanks{The authors are with the Antennas, Propagation and Micro-wave Systems (APMS) section, Department of Electronic Systems, Aalborg University, Denmark (e-mail: $\left\{\mbox{pe,stz}\right\}$@es.aau.dk).}  
\thanks{The motivation behind this work was spurred by a pilot study we carried out for Huawei Gothenburg, Sweden.} 
}

\maketitle

\begin{abstract}
Obtaining down link (DL) channel  state information  (CSI) at the base  station (BS) is challenging for  frequency-division-duplex (FDD) massive MIMO (MM) systems. Considerable  overhead  is required for DL training  and   feedback. Instead studies often assume highly correlated average spatial signal signatures (i.e. directional clusters) between FDD duplex links. These assumptions, however, only represent use of antennas with the same radiation pattern over the duplex band, leading to illumination of the same cluster.

In this paper, we investigate pattern reciprocity, over the duplex band, for practical user handsets. We first show how a population of measured contemporary phones exhibits noticeable duplex pattern divergence.
We then show measured complex pattern duplex divergence of a mock-up phone, where depolarization comes on top of gain differences.
Thus we reveal a significant but overlooked brick in DL CSI assessment for FDD MM operation. Namely, in order for the FDD MM performance studies to be realistic, practical user handsets have to be considered.
\end{abstract}

\IEEEpeerreviewmaketitle

\section{Introduction} 
\label{sec:introduction}

Multiple-input-multiple-output (MIMO) antenna systems have been introduced in (3G) 4G systems \cite{ref:ETSI 4G}, for increasing the capacity.
Further capacity enhancement is achieved by employing Massive MIMO (MM), involving more antennas in the communication process, which is a key technology for 5G systems. For time-division-duplex (TDD), up-link (UL) and down-link (DL) channel state information (CSI) has proven to have sufficient coherence for successful deployment of MM DL beam pointing algorithms based only on UL CSI. 
Frequency-division-duplex (FDD) MM operation, on the other hand, is specified for many licensed operation bands \cite{ref: 3GGP E-UTRA oper. bands}. Thus porting new 5G system into these bands imposes FDD operation.

\subsection{An inherent assumption of FDD Massive MIMO} 

FDD MM studies generally recognize that instant CSI reciprocity between UL and DL does not readily apply or does not exist  \cite{ref:FDD Massive MIMO via UL/DL Channel Covariance Extrapolation,ref: Directional training for FDD MM,ref: TDD vs FDD MM  - what measurem. say,ref: On the directional reciprocity of uplink and downlink channels,ref: FDD MM DL spatial ch est,ref:FDD MM burst CSI,ref: Towards practical FDD MM}. However, in average sense some spatio-frequency correlation is assumed\footnote{previous MM channel sounding campaigns
\cite{ref: On the directional reciprocity of uplink and downlink channels, ref: TDD vs FDD MM  - what measurem. say, ref: Channel Extrapolation for FDD MM, ref: Directional training for FDD MM,ref: Towards practical FDD MM}, 
have used simple UE antennas which can impose the same spatial duplex illumination} between UL and DL, i.e. the duplex separation is within the coherence angles and bandwidth of the channel \cite{ref:FDD Massive MIMO via UL/DL Channel Covariance Extrapolation,ref: Directional training for FDD MM,ref: TDD vs FDD MM  - what measurem. say,ref: On the directional reciprocity of uplink and downlink channels,ref: FDD MM DL spatial ch est,ref:FDD MM burst CSI,ref: Towards practical FDD MM}. 

This assumption means identity of the average UL and DL scattering cluster directions, which facilities relaxed CSI loop-back.
However, if the radiation patterns of the UE antenna at the frequency bands of UL and of DL are diverging, then there is a good chance for significant different illumination of (or at least part of) scattering clusters. Since this inherent pattern reciprocity assumption is impacting average directional responses and thus extent of needed MM DL CSI to obtain optimal beam directions, it is crucial to assess its validity. The latter is the issue addressed in this paper.

\subsection{Duplex pattern reciprocity}
From both antenna and propagation perspective, duplex pattern reciprocity is questionable if it can be generalized, even in the average sense - and this paper fills this gap with practical knowledge. While pattern reciprocity is achievable at base-station (BS) side (free propagation conditions), the story at the user equipment (UE) handset side may be very different and is of interest in this paper. The fast change of the radiation pattern over frequency, mainly depends on the size of the effective radiating structure. I.e., small simple antennas by them selves cannot change radiation pattern fast\footnote{fundamental antenna structures with size below ca. one wave length \cite{ref:Balanis}}. Thus, variation of the radiation pattern over frequency is partly due to the antenna element and partly due to the handset body and the user presence \cite{ref:IET_SSZ_AT_EF_GFP,ref:Effects on portable antennas of the presence of a person}.

We will discuss three situations considering UE antenna duplex patterns:
\begin{itemize}
\item different antenna elements used for UL and DL.
\item multiple antennas used in (typically) DL direction (i.e. receive diversity). For SAR issues also different transmit (i.e. UL) elements could be used - even dynamically.
\item same antenna element used for UL and DL, but pattern changes noticeably over the duplex separation.
\end{itemize}
while the two first situations have obvious potential for an-reciprocal duplex patterns, the latter situation is an overlooked culprit for not meeting the aforementioned assumptions.

\begin{figure}
\centering
\includegraphics[trim={5mm 105mm 0 40mm}, scale=0.41]{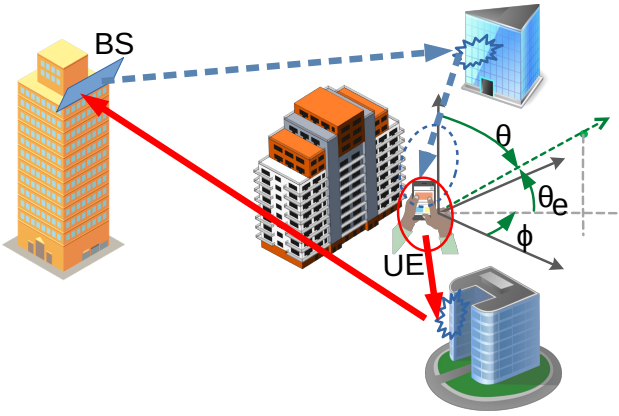} 
\caption{NLOS scenario of UE with divergent UL (solid line) and DL (dash line) patterns. Spherical coordinates are labeled with ($\theta$, $\phi$) and elevation is given by ($\theta_e$).}
\label{fig:Scenario}
\end{figure}

In line-of-sight (LOS) we essentially have an antenna pattern illumination, approaching an angular Dirac's delta function. Thus no UL vs DL angular discrimination effects will be observed. 
On the contrary, as illustrated in Fig. \ref{fig:Scenario}, pronounced angular discrimination can occur at the BS side in obstructed propagation conditions - non-line-of-sight (NLOS),  due to divergent UL vs. DL UE patterns in multi modal (i.e. several separated scattering clusters) or widely distributed scatterings scenarios. 

As shown in the paper, FDD duplex pattern an-isotrophy does exist for practical handsets and at a level strong enough to expect noticeable impact in systems studies or operation.
Throughout the paper, the normal spherical convention with azimuth $\phi$ and inclination $\theta$ - unless specifically noted, the elevation $\theta_e=\pi/2-\theta$ \cite{ref:Balanis} is used (see  Fig. \ref{fig:Scenario}). Furthermore, the variable dependencies of probability density distributions (pdf) $f$ is understood.

The  paper  is  organized  as  follows.  In Section   II   we   discuss the duplex pattern similarity metrics.   Section   III contains  analysis  of  power patterns from a wide range of measured commercial mobile phones.  Section IV  contains  analysis of measured complex patterns of antennas in a mock-up handset. Section V concludes the paper. 

\section{Antenna pattern reciprocity assessment metrics and procedures}\label{sec:Antenna pattern reciprocity metrics}
Characterizing UE antenna patterns $h$ fully, requires complex valued spherical  $\Omega=(\theta,\phi)$ and dual polarized  ($\theta$ and $\phi$ polarization)  resolution, i.e. $h(\Omega)=h_\theta(\Omega)+h_\phi(\Omega)$.

\subsection{Scalar patterns}
\label{subsec: Scalar patterns}
Often antenna performance is based only on power ($p$) measures, i.e. $p \propto |h|^2$. We use the basic scalar descriptors for UE antenna efficiency as provided in \cite{ref:IEEE_access_SSZ_GFP}: 1) the effective isotropic radiated power is  $p_{\mathrm{Tx}}|h_{\mathrm{Tx}}(\Omega)|^2=EIRP_{\Omega}=EIRP_{\theta}(\theta,\phi)+ EIRP_{\phi}(\theta,\phi)$; and 2) effective isotropic sensitivity (for a given output signal-to-noise ratio SNR) is $(p_{\mathrm{Rx}}|h_{\mathrm{Rx}}(\Omega)|^2)|_{\mathrm{SNR}}=EIS_{\Omega}=EIS_{\theta}(\theta,\phi)+ EIS_{\phi}(\theta,\phi)$.

An obvious UL vs. DL pattern comparison is via the isotropic power balance ($IPB$) as $IPB \propto EIRP(\theta,\phi) \cdot EIS(\theta,\phi)$. $IPB$ will be a perfect sphere if UL and DL patterns are identical. However, deviations from a sphere can suggest exaggerate\footnote{for example, DL pattern nulls (i.e. high $EIS$) will exhibit large relative differences but result in little absolute signal difference.} importance.

A better metric to compare patterns is a difference construct of normalized UL vs. DL isotropic patterns, as
\begin{equation}\label{eq: DIP_rel_dif_pattern}
    \Delta IP(\theta,\phi)=\left|\frac{EIRP(\theta,\phi)}{\overline{EIRP(\theta,\phi)}}  - \frac{1/EIS(\theta,\phi)}{\overline{1/EIS(\theta,\phi)}} \right| 
\end{equation}with the normalizing means taken over $\Omega_r$ (i.e. spherical region of investigation). This normalization typically gives $0.4\lesssim\overline{\Delta IP}\lesssim 1$ and is not strictly bounding the metric, but it is simple and largely suppresses variations due to absolute levels.

A very compact dual-direction metric, compares DL vs. UL selectivities at maximum (DL) power azimuth direction ($\phi_{max,DL}\forall ~\theta$ ) with its opposing (i.e. $\phi_{max,DL}+\pi$) direction, i.e. the cross link front to back ratio (FB)
\begin{multline}\label{eq: dif cross UL-DL F2B}
\Delta \mathrm{FB}_{DL,UL}=\mathrm{FB}_{DL}(\phi_{max,DL})-\mathrm{FB}_{UL}(\phi_{max,DL})\\
=p_{DL}(\phi_{max,DL})-\mathrm{max}(p_{UL}(\phi_{max,DL})) \\
+\mathrm{max}(p_{UL}(\phi_{max,DL}+\pi))-\mathrm{max}(p_{DL}(\phi_{max,DL}+\pi)) [\mathrm{dB}]
\end{multline}
This indicator shows whether the links focus on clusters in opposing directions. The maximization in  \eqref{eq: dif cross UL-DL F2B} is taken across the elevation span of $\Omega_r$.

For scalar patterns, some polarimetric insight can be gained from the shear polarimetric power decomposition, i.e. cross polarization discrimination $\mathrm{XPD}=p_\theta/p_\phi$. The XPD expresses the purity of the linear polarization, but is not so suitable\footnote{as the logarithmic mapping in the dBs is double sided open scale, it becomes ambiguous to compare e.g. (near co-polar) $\Delta_{XPD}=XPD_{DL}(+20\mathrm{dB})-XPD_{UL}(+10 \mathrm{dB})=10\mathrm{dB}$  vs. (near orthogonal) $\Delta_{XPD}=XPD_{DL}(+5\mathrm{dB})-XPD_{UL}(-5 \mathrm{dB})=10\mathrm{dB}$. Similarly for the linear singled side open scale.} for comparative purposes. Instead we use the linear polarization tilt $\psi \equiv\alpha=\arctan(E_\theta/E_\phi)=\arctan(\sqrt{p_\theta/p_\phi})$, where $\alpha$ is the help angle in the full polarization description in section \ref{subsubsec: pol. ellipse}. When XPD  is large or small we are close to linear polarization and the tilt will be close to $90^o$ or $0^o$, respectively. XPD close to 1 (0 dB) means  $\psi\approx 45^o$, but the polarization can be anything from pure linear to circular, depending on actual phase relationships. We can though make a rough polarization discrimination assessment by looking at the tilt angle difference, as 
\begin{equation}\label{eq: dif_pol_tilt}
   \Delta \psi (\theta,\phi) =|\psi_{UL}(\theta,\phi) - \psi_{DL}(\theta,\phi)| 
\end{equation}As the scalar pattern fields have no sign or phase, $\psi$ and $\Delta\psi$ are restricted to the range $0-90^o$. Polarization mismatch reveals an extra signal difference influence on top of the directional gain impacts.

\subsection{Complex patterns}
\subsubsection{Correlation} A compact descriptor for complex pattern similarity is the antenna signal $s$ cross correlation $\rho_{UL,DL}$ involving the total pattern directional complex gains and polarization state. When UL and DL patterns are assumed exposed to same cluster illuminations, polarizations mutually uncorrelated and providing equal directional illumination with a certain fixed XPD \cite[(8.4.42)-(8.4.43)]{ref:RVG+JBA} - we isolate the shear pattern correlation coefficient as

\begin{equation}\label{eq: pattern correlation coeff}
\rho_{UL,DL}=\frac{C_{UL,DL}}{\sqrt{C_{UL,UL}C_{DL,DL}}}\equiv \rho_{hj,hk} \big|_{P_S, XPD}
\end{equation}
where the antenna (pattern \textit{j} vs \textit{k} are the antenna and/or link indexes) signal ($s \propto h \cdot p_S$) correlations are 
\begin{equation}\label{eq:antenna j-k correlation}
C_{j,k}=\int p_S(\Omega) [XPD h_{j\theta}(\Omega) h^*_{k\theta}(\Omega)+ h_{j\phi}(\Omega) h^*_{k\phi}(\Omega)] d\Omega
\end{equation}and $p_S(\Omega)$ is the illuminating power distribution (i.e. scattering cluster foot print onto the antenna pattern) with assumed normalization $\int p_S(\Omega) d\Omega =1 $. Thus  $p_S(\Omega)$ can be interpreted as a probability density. Note the correlation coefficient in \eqref{eq: pattern correlation coeff} is complex. If we have diffuse scattering only, we can assume Rayleigh fading for which the often used envelope correlation is $\rho_{|s_j|,|s_k|}\approx \rho_{|s_j|^2,|s_k|^2}=|\rho_{s_j,s_k}|^2 \equiv |\rho_{hj,hk}|^2$ \cite{ref:RVG+JBA}.

To make signal decorrelation more transparent and comparable across different illumination directions and different UE handsets, we choose a fixed canonical shape cluster footprint ($p_S$) imposed onto the antennas.

Stochastic based models often favor some sort of ellipsoid cluster structure with a Laplacian ($\mathcal{L}: f_{\mathcal{L}}=\frac{1}{2b}\mathrm{exp}(-|\phi|/b)$, with $b=\sigma_\phi/\sqrt{2}$) power distribution in the (azimuth) horizontal plane ($\theta_e=0^o$) and  Laplacian or Gaussian in the vertical(elevation) plane \cite{ref:3GPP 5G cluster model}. However, our objective is exposure of DL vs UL antenna pattern differences over the sphere. Thus, we must use a pattern illumination without orientation ambiguities (e.g. over the poles).  That is, we choose a circular foot print with rotational symmetry, which after integration over elevation (around the horizontal plane), collapses to an approximate Laplacian azimuth distribution\footnote{a pilot study used a uniform distribution. The findings remains compared to the Laplacian distribution, when scaled with the spread of the distribution} (see Appendix).
When moving the center of this cluster over the sphere, we get directional sensitive duplex signal correlations. The actual processing steps are given in the Appendix. 

\subsubsection{Polarization ellipse}\label{subsubsec: pol. ellipse} For a general polarization, the  ellipse tilt $\tilde{\psi}$  is retrieved via $\mathrm{tan}(2\tilde{\psi})=\mathrm{tan }(2\alpha)\cdot \mathrm{cos}(\delta)$ and the eccentricity angle $\chi$ via $\mathrm{sin}(2\chi)=\mathrm{sin}(2\alpha)\cdot \mathrm{sin}(\delta)$. The help angle $\alpha$ is given in Section \ref{subsec: Scalar patterns}, while  $\delta=\angle E_\theta - \angle E_\phi$ is the field component phase difference \cite{ref:Balanis}. The polarization vector\footnote{with time evolution and frequency being understood} is $\vec{E}=[E_\phi ; E_\theta  ]^T$. The polarization loss factor (PLF) between antennas (and link) 1 and 2, becomes \cite{ref:Balanis}
\begin{equation}
\eta=|\vec{E_1} \cdot \vec{E_2}|^2
\end{equation}i.e. a vector projection with normalized polarization vectors $|\vec{E}|=1$. We use same convention for sense of rotation definition, no matter if antennas are in receive or transmit mode so measured patterns can be compared directly.

\subsection{Population statistics}
Global population (all phones and bands) statistics of correlations and other metrics, are taken either over the \textit{sphere} (or rather down to the measurement limit $ \Omega_r(\theta) \leq 150^o$, see next section) - or over a subset \textit{girdle} above the horizontal plane ($ 0^o\leq \Omega_r(\theta_e) \leq 30^o$).

\section{Live handset scalar patterns}\label{sec:Live handset}
\subsection{Measurement setup}\label{subsec:Measurement setup}
Details about the  handset measurement campaign, which results are used in this paper, can be found in  \cite{ref:IEEE_access_SSZ_GFP}.
Here, antenna radiation patterns in terms of EIRP and EIS of 16 contemporary mobile phones have been measured over-the-air (OTA) in 2G, 3G and 4G bands for both vertical and horizontal linear polarization.
The operating bands \cite{ref:IEEE_access_SSZ_GFP},\cite[Table 5.5-1]{ref: 3GGP E-UTRA oper. bands} we use are (UL/DL center frequencies [MHz]) : UMTS900 (897.5/942.5); UMTS2100 (1950/2140); LTE800 (847/806); LTE1800 (1747.5/1842.5) and LTE2600 (2535/2655).

\subsection{Pattern differences vs. frequency bands}

\begin{figure*}
\centering
\includegraphics[width=\textwidth]{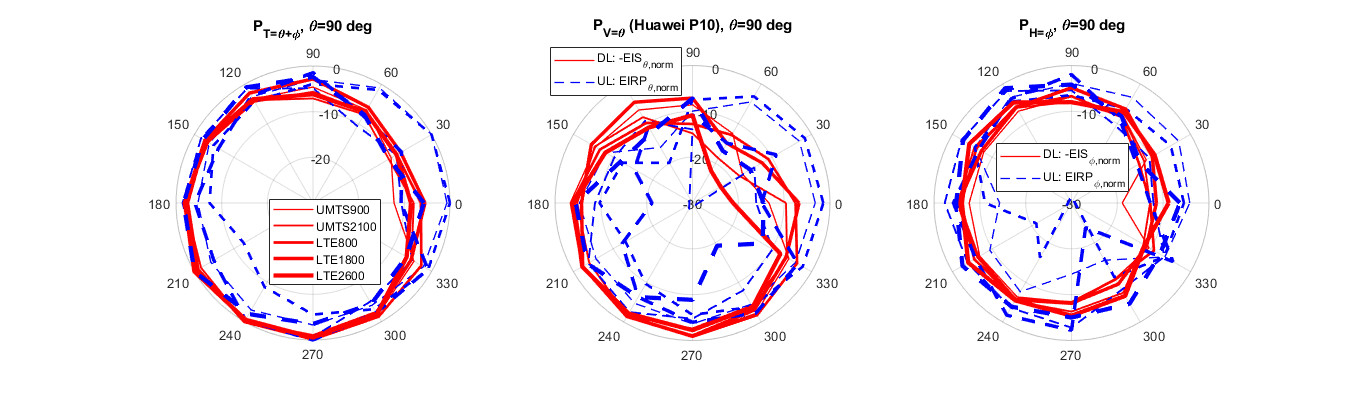} 
\caption{Huawei P10 phone, horizontal cut ($\theta=90^o$) of normalized power patterns in dB (to maximum of total power in DL or UL at $\theta=90^o$) of different operating bands (Solid: DL. Dash: UL). Left: Total power. Middle: Vertical (V, i.e. $\theta$) polarization. Right: Horizontal (H, i.e. $\phi$) polarization. }
\label{fig:Huawei_P10_power}
\end{figure*}

An example of power patterns in the horizontal plane ($\theta_e = 0^o$) is shown in Fig. \ref{fig:Huawei_P10_power} for one of the tested smartphones (using $\Delta\phi=30^o$ for both UL and DL). The DL (EIS) data have been inverted (negation in dB) so they represent proportionality with antenna gain, just as the UL (EIRP) does. It is noticed how the total power (i.e. both polarization fully exposed) is a well behaved round shape. However, even in this case there is a front-to-back (FB) ratio difference between UL and DL patterns of order 10dB for the UMTS bands. This can cause emphasis on different scattering clusters.

The individual polarization patterns exhibit significant differences between DL and UL.
This is quantified in Table \ref{tab: power metrics} for all 16 phones tested where the DL field strength ($|E| \propto \sqrt{p}$) has been linearly interpolated to give same grid resolution of $15^o$ as for the UL, to use all data available. Only two phones (Doro 7070 and Huawei P9 Lite mini) do not have any marked high ($\Delta \mathrm{FB}$) pattern divergence in any band. Those two phones also exhibit lower end standard deviations of $\Delta IP$ and $\Delta\psi$, albeit the differences here between phones is not pronounced (therefore not included in Table \ref{tab: power metrics}).

\begin{figure}
\centering
\includegraphics[trim={30mm 5mm 30mm 10mm}, scale=0.53]{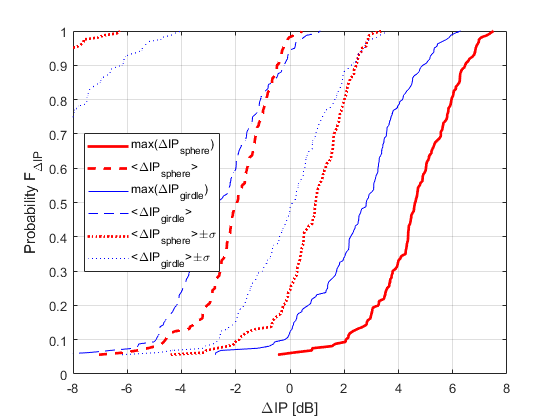} 
\caption{Cumulative probability $F_{\Delta\mathrm{IP}}$ of $\Delta\mathrm{IP}$ maximum and mean ($\pm\sigma$)  values across all phones, bands, and polarizations - for full sphere or girdle spanning $0^o\leq \theta_e \leq 30^o$}
\label{fig:DIP_stat}
\end{figure}

\begin{figure}
\centering
\includegraphics[trim={30mm 5mm 30mm 10mm}, scale=0.53]{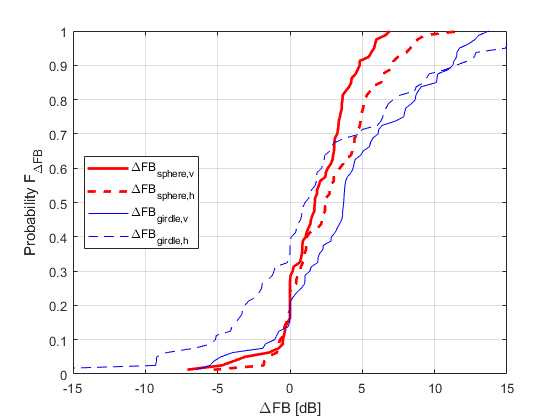} 
\caption{Cumulative probability $F_{\Delta\mathrm{FB}}$ of $\Delta\mathrm{FB}$ for vertical(v) and horizontal(h) polarization patterns,  across all phones, and bands - for full sphere or girdle spanning $0^o\leq \theta_e \leq 30^o$ }
\label{fig:DFB_stat}
\end{figure}

An example of $IPB$ for the horizontal plane is presented in Fig. \ref{fig:iPhoneX_polarization}, showing the tendency to a round  (circle) shaped response, but also significant deviations for some bands and directions. 
The over all cross link gain behaviors via $\Delta{IP}$ and $\Delta{FB}$ metrics, are shown in Fig. \ref{fig:DIP_stat} and \ref{fig:DFB_stat}.

The metric $\Delta IP$, shows population medians of $\overline{\Delta IP}+\sigma_{\Delta IP}$ around 0dB, while deciles are around the $\pm$2dB marks. So a noticeable portion of phones and bands have duplex pattern imbalances.

A very revealing and sensitive gain based metric is the $\Delta FB$, as it is a single maximization value and not a distribution over $\Omega_r$ (as $IPB$ and $\Delta IP$). The median of $\Delta\mathrm{FB}$ are modest around the 2.5dB mark. However, the upper decile for the girdle case, is around 12dB. I.e ca. 10\% of the phones and bands exhibit a strong  UL to DL divergent focus in environments simultaneously illuminating from opposing directions. This  is a potential trouble maker situation for FDD-MM CSI UL to DL translation.

\subsection{Polarization state differences}
\begin{figure*}
\centering
\includegraphics[width=\textwidth]{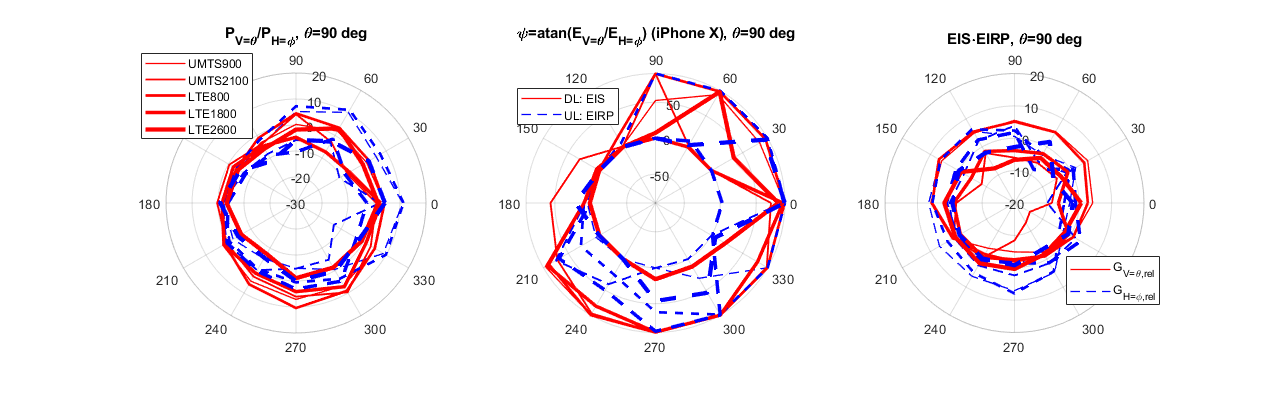} 
\caption{iPhone X, horizontal cut ($\theta=90^o$) of polarization state for  normalized power patterns (to maximum of total power in DL or UL at $\theta=90^o$) of different operating bands.  Left: XPD [dB] (Solid: DL. Broken: UL). Middle: Tilt $\psi$ [$^o$]  (Solid: DL. Dash: UL). Right: IPB for each polarization (Solid: V i.e. $\theta$. Broken: H i.e. $\phi$).}
\label{fig:iPhoneX_polarization}
\end{figure*}

The polarization state variation in the horizontal plane via the XPD and tilt $\psi$ is shown by an example in Fig. \ref{fig:iPhoneX_polarization}. As discussed in Section \ref{sec:Antenna pattern reciprocity metrics}, we resort to only using $\psi$ for analysis. Here we observe significant occurrences of UL and DL exposing orthogonal tilts, i.e. strong depolarization if close to linear polarization is present.

\begin{table}
  \caption{Cross link front-back ratios $\Delta \mathrm{FB}_{\mathrm{DL,UL}}  [\mathrm{dB}]$ ($\theta/\phi$-polarization) \eqref{eq: dif cross UL-DL F2B}, for a girdle spanning $0^o\leq \theta_e \leq 30^o$ and $\Delta\theta=\Delta\phi=15^o$. High (grayed red italics: $|\Delta \mathrm{FB}|\geq 10 \mathrm{dB}  $) and low (bold green: $|\Delta \mathrm{FB}|\leq 3 \mathrm{dB}  $) bi-modal pattern divergence.}
    \vspace{-0.75cm}
\begin{center}\label{tab: power metrics}
  \setlength\tabcolsep{2.0 pt} % default value: 6pt
  \begin{tabular}{|l?c|c|c|c|c|}
\hline
 %\multirow{1}{*}{\backslashbox{
 Phone \cite{ref:IEEE_access_SSZ_GFP} 
%{\backslashbox{Phone \cite{ref:IEEE_access_SSZ_GFP}}{Band} }
 & UMTS900  & UMTS2100 & LTE800 &  LTE1800 & LTE2600\\
     \hline
    Doro 7070 & 8.7/\tgreen{\textbf{1.9}} & \tgreen{\textbf{1.0}}/\tgreen{\textbf{-1.5}}  & 3.6/\tgreen{\textbf{0.9}}  & \tgreen{\textbf{1.0}}/\tgreen{\textbf{0.4}} &  8.4/\tgreen{\textbf{-2.0}}\\
         \hline
    Samsung G. S9 & 6.1/\tgreen{\textbf{-2.9}}&
    {\cellcolor[gray]{.8}} \tred{\textit{11.3}}/\tred{\textit{18.3}}&
    4.3/\tgreen{\textbf{1.3}}&
    \tgreen{\textbf{2.3}}/\GreyBox{\tred{\textit{12.3}}} & -5.5/6.6\\
         \hline
    Samsung G. S9+ &8.7/\tgreen{\textbf{2.8}} & 9.4/\GreyBox{\tred{\textit{10.9}}} &  3.1/\tgreen{\textbf{1.0}} & {\cellcolor[gray]{.8}} \tred{\textit{13.7}}/\tred{\textit{13.7}} & 
    4.2/9.6\\
         \hline
    Samsung G. S8 & 7.3/-3.8 & 9.1/\GreyBox{\tred{\textit{15.0}}} &   3.8/\tgreen{\textbf{0.6}} &   \tgreen{\textbf{2.9}}/7.6 & 
    \tgreen{\textbf{2.0}}/9.3  \\
         \hline
    Huawei P20 Pro &\tgreen{\textbf{0.2}}/\tgreen{\textbf{2.8}}&
    \tgreen{\textbf{-1.7}}/\GreyBox{\tred{\textit{20.8}}}& \tgreen{\textbf{0.0}}/\tgreen{\textbf{1.8}}&
    -4.8/\GreyBox{\tred{\textit{22.9}}}& 4.9/\GreyBox{\tred{\textit{14.9}}} \\
         \hline
    Nokia 7 Plus & \tgreen{\textbf{1.9}}/6.5&
    -4.0/-5.2 & 
    \GreyBox{\tred{\textit{13.1}}}/6.4 & \tgreen{\textbf{2.3}}/-6.6 &
    -6.5/9.4\\
         \hline
    iPhone 7 & 3.4/-9.3 &
    \GreyBox{\tred{\textit{11.7}}}/6.0&
    3.8/-3.1&
    \GreyBox{\tred{\textit{10.9}}}/3.1& 5.7/\tgreen{\textbf{2.4}}\\
           \hline
    iPhone 8 &3.3/-8.5&
    8.7/\tgreen{\textbf{0.1}}&
    4.0/-4.1&
    \GreyBox{\tred{\textit{10.2}}}/-3.3&
    \tgreen{\textbf{2.8}}/\tgreen{\textbf{-0.2}} \\
         \hline
    iPhone X & 3.7/\tgreen{\textbf{-2.3}} & 3.7/\tgreen{\textbf{-1.9}}&  3.9/\GreyBox{\tred{\textit{10.8}}} & \tgreen{\textbf{0.1}}/\tgreen{\textbf{1.5}} & -5.3/\tgreen{\textbf{0.7}} \\
         \hline
    iPhone 8 Plus  &6.1/\GreyBox{\tred{\textit{12.9}}}& 8.9/\tgreen{\textbf{-1.1}}& {\cellcolor[gray]{.8}}\tred{\textit{12.8}}/\tred{\textit{11.4}}&
    6.4/-5.1& 
    5.5/3.8 \\
         \hline
    Sony Xperia XA2 & \tgreen{\textbf{1.4}}/\tgreen{\textbf{2.2}} &  
    7.9/8.7 &
    \GreyBox{\tred{\textit{11.5}}}/\tgreen{\textbf{2.8}} &
    3.6/\tgreen{\textbf{1.3}} & \GreyBox{\tred{\textit{10.1}}}/ 6.3 \\
         \hline
    OnePlus 6 &7.9/\tgreen{\textbf{-0.2}}&
    3.6/\tgreen{\textbf{0.8}}& \GreyBox{\tred{\textit{11.3}}}/-3.5&
    \tgreen{\textbf{1.0}}/\GreyBox{\tred{\textit{-18.4}}}&\tgreen{\textbf{0.8}}/\tgreen{\textbf{2.4}}\\
         \hline
    Huawei P10 lite &5.6/\tgreen{\textbf{1.0}}&
    \tgreen{\textbf{0.5}}/7.1&
    8.0/\tgreen{\textbf{1.7}}& \tgreen{\textbf{-1.1}}/-9.2&
    \GreyBox{\tred{\textit{13.8}}}/6.8 \\
         \hline
    Huawei P9 lite m. &3.8/\tgreen{\textbf{1.8}}& \tgreen{\textbf{-1.9}}/\tgreen{\textbf{-1.2}}& \tgreen{\textbf{-0.8}}/-4.0&
    \tgreen{\textbf{-0.9}}/\tgreen{\textbf{2.3}}&
    3.1/5.0 \\
         \hline
    iPhone Xs Max &3.7/8.6&
    \tgreen{\textbf{1.5}}/\tgreen{\textbf{0.5}}&
    \GreyBox{\tred{\textit{10.2}}}/4.9& 4.5/\GreyBox{\tred{\textit{12.8}}}&
    \tgreen{\textbf{-0.1}}/-5.4  \\
         \hline
    Huawei P10 &\GreyBox{\tred{\textit{11.5}}}/-9.2  &
    5.1/-3.3&
    \GreyBox{\tred{\textit{12.3}}}/\tgreen{\textbf{-2.1}}&
    \tgreen{\textbf{1.9}}/\tgreen{\textbf{-1.0}}  &
    4.4/\tgreen{\textbf{0.8}} \\
    \hline
  \end{tabular}
  \end{center}
\end{table}

\begin{figure}
\centering
\includegraphics[trim={30mm 5mm 30mm 10mm}, scale=0.53]{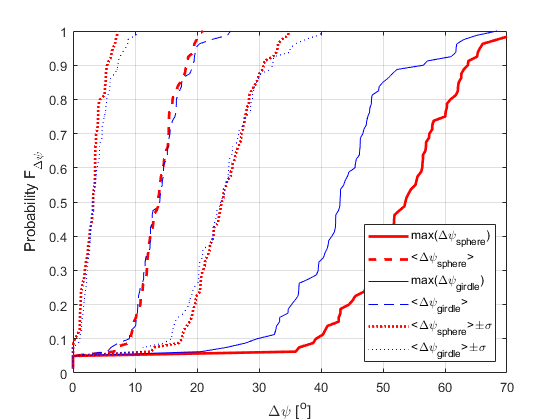} 
\caption{Cumulative probability $F_{\mathrm{\Delta\psi}}$ of $\Delta\psi$ for vertical(v) and horizontal(h) polarization patterns,  across all phones and bands - for full sphere or girdle spanning $0^o\leq \theta_e \leq 30^o$ }
\label{fig:DPsi_stat}
\end{figure}

The overall cross link polarization matching aspect is shown in the $\Delta\psi$ statistics in Fig. \ref{fig:DPsi_stat}.
While the median of  $\overline{\Delta\psi}+\sigma_{\Delta \psi}$ in the girdle case is around $24^o$,  the upper decile of the maximum $\Delta\psi$ approaches $56^o$. In linear polarization conditions, this poses corresponding matching losses of order 0.8 dB and 5 dB, respectively. Thus the maximum $\Delta\psi$ represents noticeable diverging CSI for the duplex links. 

\subsection{Total signal impact}\label{sec:total signal impact}
Despite the lack of phase information in the data from \cite{ref:IEEE_access_SSZ_GFP} we use a correlation analysis which will include directional gain and to some extent the polarimetric state as well (when links are close to linear polarization, but differ in tilt). The result will be an optimistic UL-DL signal correlation (i.e. higher or equal to that of complex patterns). Consequently, we can use these correlations for worst case assessment.

\subsubsection{Cross polarization impact}
Very large or very small\footnote{very small XPD means near all power is cross coupled into the orthogonal polarization. The mathematical impact in \eqref{eq: pattern correlation coeff} is similar to that of large XPD - however unlikely to occur in normal propagation environments.} XPD will emphasize a single polarization - and as seen from Fig. \ref{fig:Huawei_P10_power} the individual polarization tend to have larger UL vs. DL pattern differences than for the total power. Consequently we can expect larger decorrelation for very large (or very small) XPD, as encountered in more open (e.g. rural) environments. In dense scatter (e.g. urban) environments, the XPD will be closer to unity. For this reason we have used both XPD = 3 dB and XPD = 12 dB for our analysis, albeit only the 3dB case is illustrated in following sections as our numerical results reveal only minor differences to the 12 dB case.

\subsubsection{Duplex correlations}
3GPP suggests cluster azimuth spreads ranging $15^o-22^o$ in NLOS \cite[Table 7.3-6]{ref:3GPP 5G cluster model}. We choose to work with $\sigma_{rel}=15$, which gives a cluster footprint diameter of ca. $60^o$ for use with \eqref{eq: pattern correlation coeff}. The correlation statistics of the individual phones is given in Table \ref{tab: corr abs patterns}. It is noticed in general the mean correlations are very high (except a few cases indicated with red italics). The minimum correlations are typically in the 0.7 to 0.8 range, but a few cases have high minimums and compressing the distribution to the high range (indicated with bold green).

\begin{figure}
\centering
\includegraphics[trim={3mm 1.2cm  11mm 0mm},clip, scale=0.55]{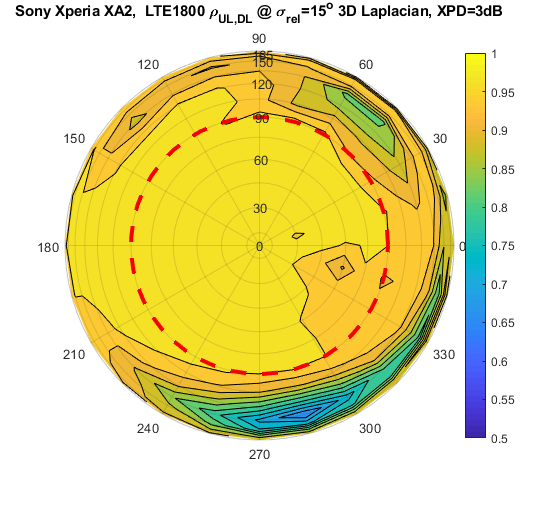}  %[scale=0.6]{Sony_Xperia_XA2_LTE1800_srel15_XPD3.png} 
\caption{High correlation LTE1800 case (XPD is 3dB) of scalar duplex patterns in \eqref{eq: pattern correlation coeff}. Dash line is horizontal plane ($\theta=90^o$). }
\label{fig:Sony Xperia_cc}
\end{figure}

\begin{figure}
\centering
\includegraphics[trim={3mm 1.2cm  11mm 0mm},clip, scale=0.55]{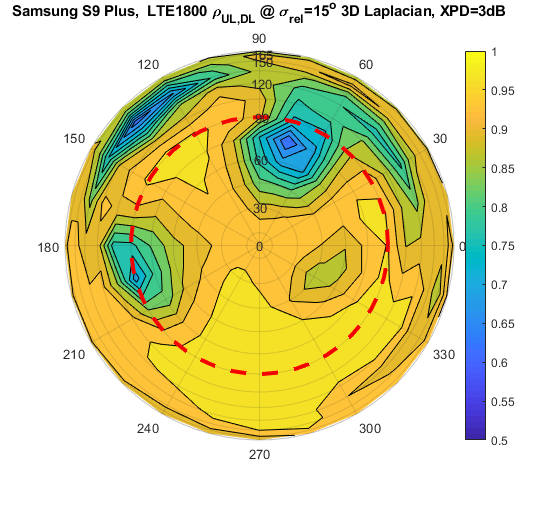} 
\caption{Bi-modal `nulls' correlation LTE1800 case (XPD is 3dB) of scalar duplex patterns in \eqref{eq: pattern correlation coeff}. Dash line is horizontal plane  ($\theta=90^o$). }
\label{fig:Samsung S9_cc}
\end{figure}

Fig. \ref{fig:Sony Xperia_cc} shows a good cross link correlation case in the LTE1800 band, with the upper hemisphere nearly all highly correlated UL and DL. This is an example of a good phone for FDD-MM operation in this band. Fig. \ref{fig:Samsung S9_cc} shows a potential problematic case in the LTE1800 band, with two distinct de-correlation `null' directions in the important range just above the horizontal plane. This Samsung S9+ phone also has high $\Delta\mathrm{FB}$ entries for LTE1800 in Table \ref{tab: power metrics} - for both polarizations. Looking into the data, the $\phi_{max,DL}$ for vertical and horizontal polarizations are $180^o$ and $90^o$, respectively, i.e. the vicinity of the two correlations nulls. This indicates $\Delta \mathrm{FB}$ usefulness as compact metric for multi cluster cross-link signal condition assessment.

\begin{table}
  \caption{Minimum (bold green: $\mathrm{min}(\rho) \geq 0.9$ ) and mean (grayed red italics:  $\overline{\rho}\leq 0.9$ ) cross link correlations $\rho_{\mathrm{UL,DL}}$ \eqref{eq: pattern correlation coeff} of scalar antenna patterns $|h|$, 3D Laplacian cluster \eqref{eq:pdf_3D_L} with $\sigma_{\mathrm{rel}}=15$ and XPD $=3\mathrm{dB}$. Data for a girdle $0^o\leq \theta_e \leq 30^o$ and $\Delta\theta=\Delta\phi=15^o$.}
  \vspace{-0.75cm}
\begin{center}\label{tab: corr abs patterns}
  \setlength\tabcolsep{1.8pt} % default value: 6pt
  \begin{tabular}{|l?c|c?c|c?c|c?c|c?c|c|}
\hline

  & \multicolumn{2}{c?}{UMTS900}  & \multicolumn{2}{c?}{UMTS2100} & \multicolumn{2}{c?}{LTE800} & \multicolumn{2}{c?}{LTE1800} & \multicolumn{2}{c|}{LTE2600}\\
 \cline{2-11}

Phone \cite{ref:IEEE_access_SSZ_GFP}
 & min &  $\overline{\rho}$ &  min &  $\overline{\rho}$ & min &  $\overline{\rho}$ &  min &  $\overline{\rho}$& min &  $\overline{\rho}$ 
\\
     \hline
    Doro 7070 & 0.73 & 0.96 & 0.84 & 0.93  & 0.81 & 0.96 &  0.83 & 0.96 &  0.75 & 0.93 \\
         \hline
    Samsung G. S9 &  0.78 &0.96& 0.73&0.93& 0.82&0.95&
    0.60&0.92&
    0.77&0.95\\
         \hline
    Samsung G. S9+ &  0.76 &0.94 & 0.66&0.94& 0.79&0.95 & 0.62&0.91& 0.81&0.94\\ %\tred{data same as S) ??? = porbelm solevd it was S9 pickign up S9+ data}\\
         \hline
    Samsung G. S8 &0.85&0.96& 0.70&0.92& 0.84&0.96& 0.59&0.92& 0.83&0.96\\
         \hline
    Huawei P20 Pro  &0.75&0.94& 0.70&0.92& 0.71&0.94& 0.63&0.90& 0.72&0.93\\
         \hline
    Nokia 7 Plus &0.79&0.94& 0.73&0.94& 0.70&0.90& 0.75&0.94& 0.65&0.91\\
         \hline
    iPhone 7 &0.89&0.95& 0.62&{\cellcolor[gray]{.8}}\tred{\textit{0.84}}& \tgreen{\textbf{0.92}}&0.98& 0.69&{\cellcolor[gray]{.8}}\tred{\textit{0.89}}& 0.82&0.94\\
           \hline
    iPhone 8 & 0.89&0.96& 0.67&{\cellcolor[gray]{.8}}\tred{\textit{0.86}}& \tgreen{\textbf{0.92}}&0.98& 0.64&{\cellcolor[gray]{.8}}\tred{\textit{0.89}}& 0.87&0.96\\
         \hline
    iPhone X & 0.84 & 0.96 & 
    0.72&0.93 & 
    \tgreen{\textbf{0.93}}&0.98& 0.86&0.97& 0.80&0.97\\
         \hline
    iPhone 8 Plus &0.46&{\cellcolor[gray]{.8}}\tred{\textit{0.86}}& 0.79&0.93& 0.55&0.91& 0.70&0.92& 0.81&0.96\\
         \hline
    Sony Xperia XA2 &  0.83&0.96& 0.76&0.93& 0.52&{\cellcolor[gray]{.8}}\tred{\textit{0.88}}& \tgreen{\textbf{0.90}}&0.97& 0.68&0.90 \\
         \hline
    OnePlus 6 &0.69&0.92& 0.78&0.93& 0.79&0.95& 0.74&0.92& \tgreen{\textbf{0.92}}&0.99\\
         \hline
    Huawei P10 lite &0.74&0.93& 0.70&0.93& 0.77&0.93& 0.75&0.94& 0.68&0.93\\
         \hline
    Huawei P9 lite m. &0.87&0.97& 0.73&0.94& \tgreen{\textbf{0.94}}&0.99& 0.78&0.98& 0.79&0.94\\
         \hline
    iPhone Xs Max &0.78&0.91& 0.80&0.94& 0.79&0.92& 0.77&0.93& 0.73&0.93 \\
         \hline
    Huawei P10 &0.81&0.95& 0.80&0.95& 0.77&0.94& 0.72&0.92& 0.73&0.93\\
    \hline
  \end{tabular}
  \end{center}

\end{table}

\begin{figure}
\centering
\includegraphics[trim={30mm 5mm 30mm 10mm}, scale=0.53]{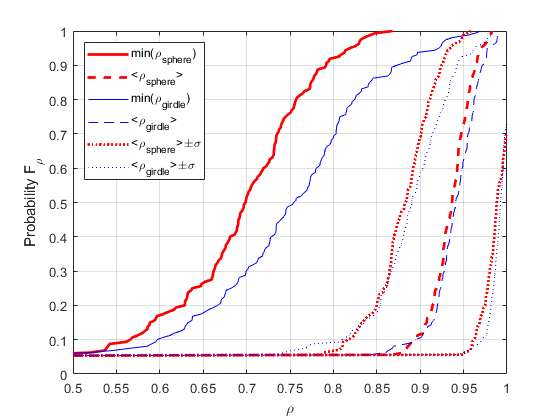} 
\caption{Cumulative probability $F_{\rho}$ of pattern correlations $\rho$  of UL vs DL scalar patterns,  across all phones and bands - for full sphere or girdle spanning $0^o\leq \theta_e \leq 30^o$}
\label{fig:cc_stat}
\end{figure}

The overall correlation statistics are shown in Fig. \ref{fig:cc_stat}.
The median of $\rho-\sigma_\rho$ is around 0.9 for the  girdle case. The corresponding median for the minimum $\rho$ is 0.75. This indicates significant duplex envelope signal de-correlations of a phone population to impact a general FDD-MM operation.

\section{Mock-up handset complex patterns}\label{sec:Mock-up}
\subsection{Measurement setup}\label{subsec:mock-up meas. setup}

We have performed high density ($\Delta\phi=\Delta\theta=3^o$) measurements of the two element antenna array mock-up phone (in the presence of a right hand phantom \cite[Fig. 3(b)]{ref:IET_SSZ_AT_EF_GFP}) - in the same bands as \cite{ref:IEEE_access_SSZ_GFP}, i.e. using the center link frequencies given in Section \ref{subsec:Measurement setup}. Antenna 1 is located at the top left corner of the mock-up phone, while antenna 2 at the bottom right corner as shown in  \cite[Fig. 3(b)]{ref:IET_SSZ_AT_EF_GFP}. Furthermore, to test effect of reduced duplex spacing, two extra frequencies are measured at the LTE2600 band, yielding three modified duplex cases, a: ($\overline{f_{UL}}+20\mathrm{MHz}$), b: ($\overline{f_{UL}}+20\mathrm{MHz}$ and $\overline{f_{DL}} - 20\mathrm{MHz}$) and c: ($\overline{f_{DL}}-20\mathrm{MHz}$).

\subsection{Comparing scalar vs complex pattern differences} \label{subsec: Compare scalar vs complex}
\subsubsection{Correlation comparisons}\label{subsubsec: corr. comparison}
\begin{figure}
\centering
\includegraphics[trim={3mm 1.2cm  11mm 0mm},clip, scale=0.55]{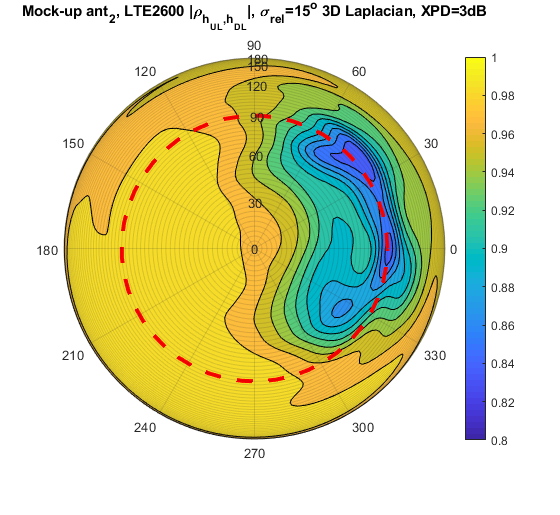} 
\caption{Complex duplex correlation \eqref{eq: pattern correlation coeff} of antenna 2 at LTE2600.  Broken line is horizontal plane  }
\label{fig:complex_cc}
\end{figure}

\begin{figure}
\centering
\includegraphics[trim={3mm 1.2cm  11mm 0mm},clip, scale=0.55]{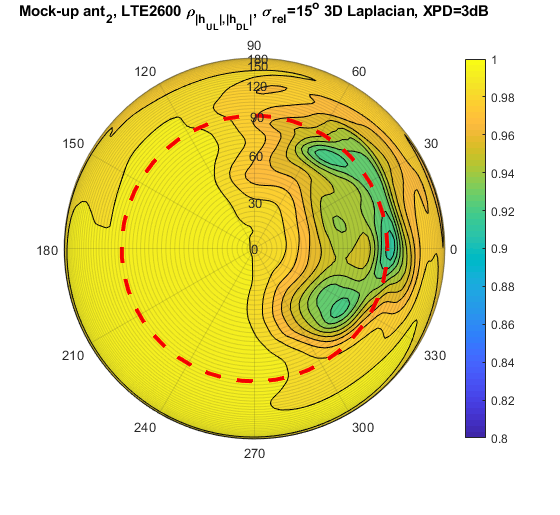} 
\caption{Duplex correlation of scalar patterns in \eqref{eq: pattern correlation coeff} for antenna 2 at LTE2600. Broken line is horizontal plane.}
\label{fig:complex_ccabs}
\end{figure}

As expected, one can see in Fig. \ref{fig:complex_cc} that the complex pattern correlation $|\rho_{h_1,h_2}|$ has larger deviations than the corresponding scalar pattern correlation $\rho_{|h_1|,|h_2|}$ shown in Fig. \ref{fig:complex_ccabs}. But even this co-element duplex comparison, shows noticeable pattern divergence relevant to be factored in for FDD MM operation studies.

\begin{figure}
\centering
\includegraphics[trim={15mm 1mm  11mm 0mm},clip, scale=0.6]{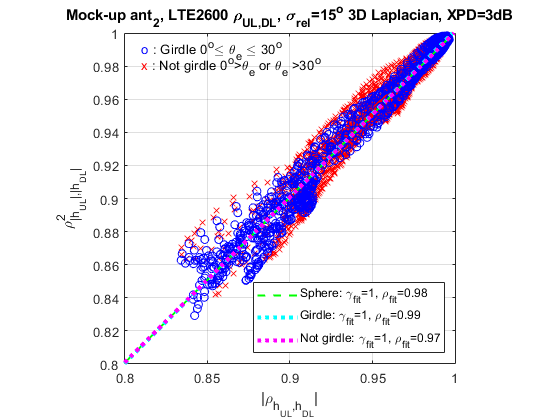} 
\caption{Comparison of magnitude of complex correlations vs. square scalar pattern correlations (linear regression slope $\gamma_{fit}$ and fitting correlation $\rho_{fit}$), for full sphere or a girdle  $0^o\geq \theta_e \geq 30^o$ - or the region outside girdle.} 
\label{fig:compare_cc_vs_ccabs}
\end{figure}

As seen from the single antenna operation example in Fig. \ref{fig:compare_cc_vs_ccabs} the square of the scalar pattern correlation approximates the magnitude of the complex correlation ($\rho^2_{|h1|,|h2|}\approx |\rho_{h1,h2}|$).

In the Rayleigh fading case we have envelope signal correlation $\rho_{|s1|,|s1|} \approx |\rho_{s1,s1}|^2$ \cite{ref:RVG+JBA}. Consequently, estimating NLOS envelope signal correlation from the scalar antenna patterns due to  \eqref{eq: pattern correlation coeff}, can give  
\begin{equation} \label{eq: cc quadruple rule}
\rho_{|s1|,|s2|} \approx |\rho_{s1,s2}|^2 \equiv |\rho_{h1,h2}|^2 \approx \rho_{|h1|,|h2|}^4
\end{equation}
i.e. a quadruple power relationship. Thus, seemingly high correlations in the scalar pattern analysis of single antenna operation should be evaluated with this relation in mind.

From our modified duplex band investigations, it follows that reducing LTE2600 duplex separation from 120 to 80MHz (LTE2600b) does not provide large rise in duplex pattern correlation. A further shrinking of the duplex separation below 1\% (20MHz), (LTE2600a and LTE2600b) is needed to bring near-to full correlation.

\subsubsection{Impact of cluster foot print size}\label{subsubsec: Cluster footpriont size}
Larger footprints give larger averaging area, so local directional `dents' in the correlation sphere will be smoothed out. Smaller footprints will exhibit larger deviations - until a point where the foot print is so small that it does not induce any significant difference illumination over the patterns (like in LOS). In this case there is no decorrelation as such to be exposed.

\begin{figure}
\centering
\includegraphics[trim={0mm 0mm  0mm 0mm},clip, scale=0.6]{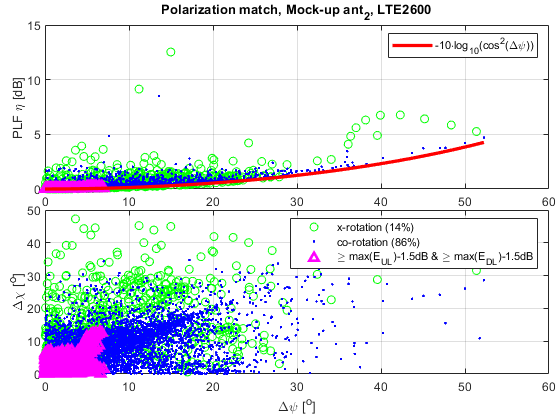} 
\caption{Polarization loss $\eta$ (top) and eccentricity  difference $\Delta\chi=|\chi_{UL}-\chi_{DL}|$ (bottom) between UL and DL of antenna 1- vs. the linear polarization tilt difference $\Delta\psi$. Circular polarization state is indicated with  co- (dot) vs cross-rotation (circle) points.}
\label{fig:pol_state2}
\end{figure}

\begin{figure}
\centering
\includegraphics[scale=0.6]{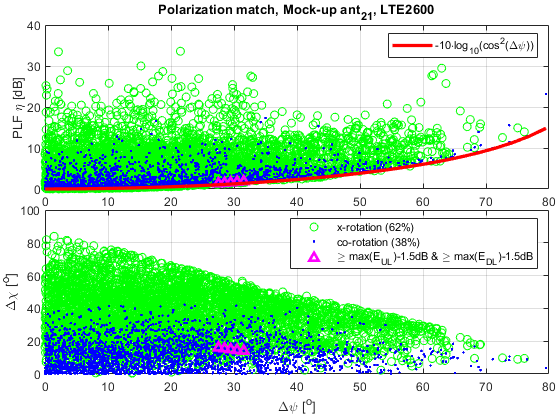} 
\caption{Polarization loss $\eta$ (top) and eccentricity  difference $\Delta\chi=|\chi_{UL}-\chi_{DL}|$ (bottom) between UL on antenna 1 and DL on antenna 2 - vs. the linear polarization tilt difference $\Delta\psi$. Circular polarization state is indicated with  co- (dot) vs cross-rotation (circle) points.} 
\label{fig:pol_state21}
\end{figure}

\subsubsection{Polarisation state}\label{subsubsec:Pol. state}
To assess the usefulness of \eqref{eq: dif cross UL-DL F2B}, \eqref{eq: dif_pol_tilt} for scalar patterns, we compare with the full polarisation state conditions. Notice the apparent upper bound on $\Delta\chi=|\chi_{UL}-\chi_{DL}|$ vs. $\Delta\psi$ \eqref{eq: dif_pol_tilt} in Fig. \ref{fig:pol_state21}.  A linear regression between $\Delta\psi$ and $\mathrm{max}(\Delta\chi)$ gives mean fitting slopes $\overline{\gamma_{fit}}=-0.75$ and mean fitting correlations of $\overline{\rho_{fit}}=0.83$ (across all bands and antenna combinations). For the most clear 'upper bound' cases, it appears  
\begin{equation}\label{eq: max(Dchi) vs Dpsi}
\mathrm{max}(\Delta\chi)|_{\Delta\psi }\approx \mathrm{max}(\Delta\psi) - \Delta\psi
\end{equation}
As we use the linear polarization tilt in \eqref{eq: dif_pol_tilt}, $\psi\equiv\alpha$. The abscissa crossing leads to linear cross polarization while the ordinate crossing leads to divergent polarization form (linear vs circular). 

For single duplex antenna element operation as in Fig. \ref{fig:pol_state2}, a few $\Delta\chi$ values above this `bound' appear. Thus \eqref{eq: max(Dchi) vs Dpsi} should be regarded as a rough rule of thumb instead of a hard limit.

\subsection{Handset multi antenna aspects}
\subsubsection{Different UL and DL elements} 
\begin{figure}
\centering
\includegraphics[trim={3mm 1.2cm  11mm 0mm},clip, scale=0.55]{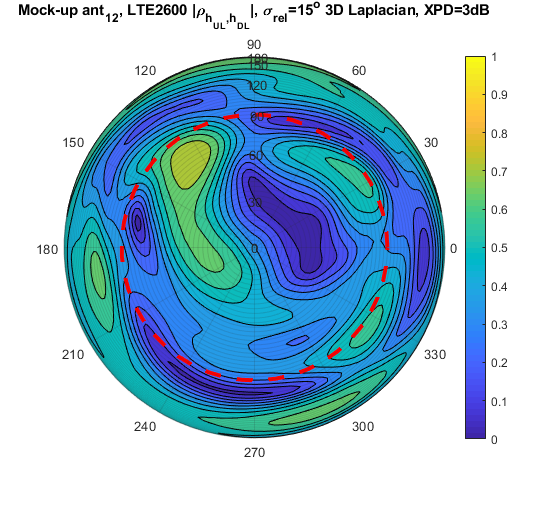} 
\caption{Complex duplex correlation \eqref{eq: pattern correlation coeff} of antenna 1 at UL LTE2600 and antenna 2 at DL LTE2600. Dash line is horizontal plane  }
\label{fig:complex_cc_12}
\end{figure}

\begin{figure}
\centering
\includegraphics[trim={3mm 1.2cm  11mm 0mm},clip, scale=0.55]{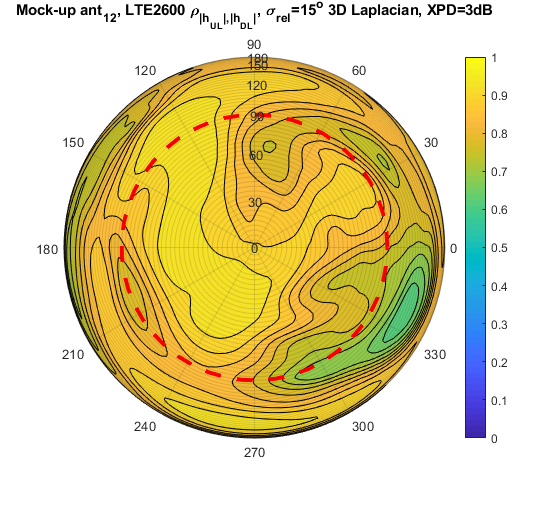} 
\caption{Duplex correlation of scalar patterns in \eqref{eq: pattern correlation coeff} of antenna 1 at UL LTE2600 and antenna 2 at DL LTE2600. Dash line is horizontal plane.}
\label{fig:complex_ccabs_12}
\end{figure}

Using antenna 1 on UL and antenna 2 on DL (or vise versa), we see in Fig. \ref{fig:complex_cc_12} a dramatic impact of phase center translation via antenna separation (this is not visible in scalar pattern correlations in Fig. \ref{fig:complex_ccabs_12}), i.e. the quadruple rule of \eqref{eq: cc quadruple rule} does not hold here.


\subsubsection{Emulated Receive (DL) diversity}
To mimic the operating condition for the live handsets in the test setup, we expose the mock-up antennas to a single direction (i.e. point source) at the time. Then a selection combiner (SC) operation at the handset gives $E_\mathrm{SC,pt}(\theta,\phi)=\mathrm{max} [  E_{\theta_1}, E_{\theta_2}] + \mathrm{max} [  E_{\phi_1}, E_{\phi_2}]$ \cite{ref:RVG+JBA}- with subscript $1$ or $2$ identifying the two antennas (and/or duplex links). A maximum ratio combining (MRC) operation will result in $E_\mathrm{MRC,pt}(\theta,\phi)\propto\sqrt{|E_{\theta_1}|^2+|E_{\theta_2}|^2} + \sqrt{|E_{\phi_1}|^2+ |E_{\phi_2}|^2}$, i.e. a scalar valued response due to the ideal phase compensation assumed in the MRC weights \cite{ref:RVG+JBA}.

\begin{figure}
\centering
\includegraphics[trim={3mm 1.2cm  11mm 0mm},clip, scale=0.55]{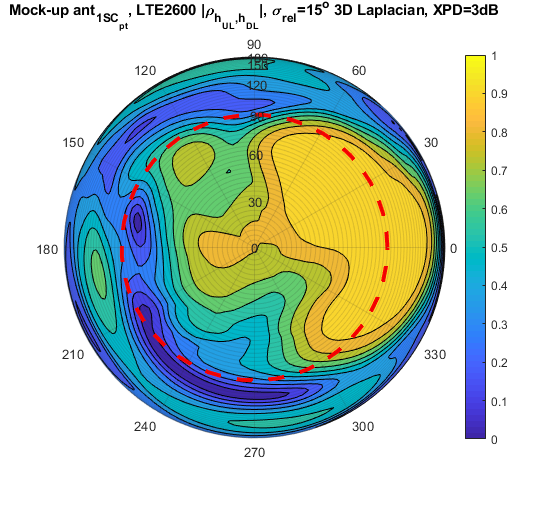} 
\caption{Complex duplex correlation \eqref{eq: pattern correlation coeff} of antenna 1 at UL LTE2600 and SC at DL LTE2600.  Dash line is horizontal plane  }
\label{fig:complex_cc_1SC}
\end{figure}

\begin{figure}
\centering
\includegraphics[trim={3mm 1.2cm  11mm 0mm},clip, scale=0.55]{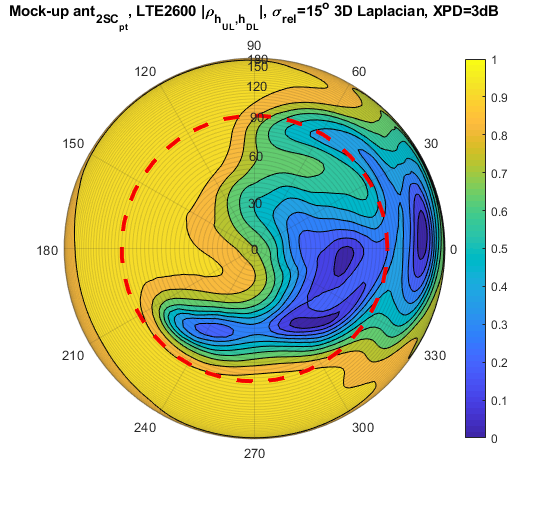} 
\caption{Complex duplex correlation \eqref{eq: pattern correlation coeff} of antenna 2 at UL LTE2600 and SC at DL LTE2600.  Dash line is horizontal plane  }
\label{fig:complex_cc_2SC}
\end{figure}

Fig. \ref{fig:complex_cc_1SC} shows emulated receive SC at DL and using antenna 1 for UL transmission. Notice the correlation breaking up into two majors regions. At left bottom same structure is visible as in Fig. \ref{fig:complex_cc_12}. That is, here antenna 2 is selected at DL (i.e. different element than UL). The opposing region with high correlation is when antenna 1 is selected as DL, i.e. same UL and DL element used. Comparing with Fig. \ref{fig:complex_cc_2SC}, as expected we see the high vs. low correlation regions are switched, when using antenna 2 for UL instead.

Even though the SC correlations show some regions of co-element duplex operation, the total correlation coverage is so low that we also in gross sense can regard the effective  duplex patterns for uncorrelated. The scalar pattern correlations look much like Fig.\ref{fig:complex_ccabs_12}  and are omitted here.

The MRC case is different as it removes the phase relation associated to each antenna. The complex correlation results looks much like that of Fig.\ref{fig:complex_cc_12} and is omitted here. Like the SC case, the scalar pattern correlations are high and can be difficult to distinguish from some single element operation correlations cases.

\section{Conclusions}
The paper shows that a common assumption of spatial coherence (i.e. average directional commonality) between FDD duplex links, is a significant over-simplification when considering practical UE handsets.
The handset population statistics from power patterns of live phones, show a noticeable fraction of cases with duplex pattern divergence to a degree it is expected to have an UL to DL CSI translation impact.

Despite the live phone data are not able directly to tell us whether single element operations is in place at some bands - our complex mock-up phone analyses shows significant duplex differences can occur even for co-element duplex operation.
This is an important duplex pattern divergence factor, which omission will lead to incorrect assumptions for the average directional reciprocity at the BS.
Furthermore, the complex data from a mock-up phone suggests a simple quadruped rule (assuming single element operation) between scalar pattern correlations and that expected of signal envelopes. Thus in this case the scalar data analysis can be ported to expected (upper bound) signal impact.

We have also emulated effects of different UL and DL elements, or handset receive diversity operation. Both cases expose very low cross-link correlations, so the duplex patterns can be regarded practically orthogonal \cite{ref:RVG+JBA}. This is yet an uncertainty factor for FDD MM coherence assumptions.

From the range of scalar pattern correlations we have a suggestion\footnote{Third party test houses do not have readily access to internal handset workings (information or control of multi antenna operation). This is a privilege of manufacturers in dedicated test phones.} of what to expect from a live phone: A minimum $\rho_{||}$ above 0.9-0.95 range suggests single element operation, where average $\rho_{||}$ below 0.9 range suggests different Tx and Rx elements (and possibly some multi antenna operation on top). This latter case can be regraded as exhibiting  fully uncorrelated duplex links. For some bands and some phones (Phone 7, 8, 8 Plus and Sony Experia XA2 particularly at UMTS2100 and LTE1800) is observed possibility for having the second scenario (different UL and DL antennas).

The findings in this paper are of a great interest for:
\begin{itemize}
\item system designers to factor in a phone population with duplex divergent patterns, in their FDD MM beam creation.
\item networks to rate phones with cross-link pattern fidelity according to expected success for FDD MM operation.
\item handset antenna designers to create a new cross-link pattern fidelity design criteria (dedicated FDD-MM phones).
\end{itemize}

To relieve cross-link pattern differences of single antenna operation, our mock-up phone measurements indicated we need to go down to order  1\%  duplex separation. So small duplex separation is not supported in current FDD license bands \cite{ref: 3GGP E-UTRA oper. bands}.

\begin{figure}
\centering
\includegraphics[trim={6mm 3mm  6mm 3mm},clip,scale=0.55]{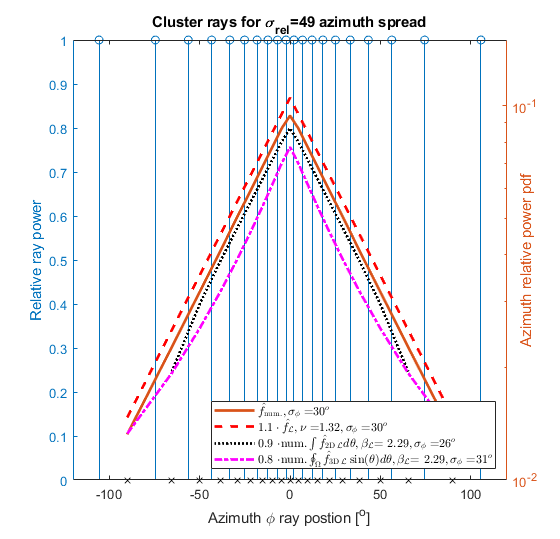} 
\caption{Azimuth power density probabilities for hemispherical footprint ($\sigma_{\mathrm{rel}}=49$). Left: (top '$\circ$') discrete ray powers, (bottom '$x$') midpoint interval positions. Right: truncated Laplacian (solid: \eqref{eq:pdf_f_num}, broken \eqref{eq:pdf_f_L}) and integrated 2D approximation (dotted: \eqref{eq:pdf_2D_L}) and correspondingly in 3D (dash-dot: \eqref{eq:pdf_3D_L}).  }
\label{fig:circ_footprint}
\end{figure}

\appendix[Circular cluster footprint]
The Laplacian azimuthal cluster ray definition in \cite{ref:3GPP 5G cluster model}, spaces $N=20$ uniform (relative) power components $p_n$ at varying angular density. For $\sigma_{\mathrm{nom}}=1^o$ nominal azimuth spread, the ray locations $\phi_n$ span between $\pm \text{max}(\phi_{\mathrm{nom}}) = \pm 2.1551^o$.
We transform to a varying component power representation (i.e. at $M=N-1$ angular midpoints $\phi_{m}=\frac{1}{2}(\phi_{n+1}+\phi_n)$ for numerical evaluation in the following), i.e.
\begin{equation}\label{eq:pdf_f_num}
\hat{f}_{num}\propto g_{num}(\phi_m)=\frac{1}{\Delta\phi_n}
\end{equation}
with $\Delta\phi_n=\phi_{n+1}-\phi_n$.  The azimuth spread scales as $\sigma_\phi=\sigma_{\mathrm{rel}}\sigma_{\mathrm{nom}}$, with relative spread $\sigma_{\mathrm{rel}}$.
The corresponding truncated continuous form Laplacian is
\begin{equation}\label{eq:pdf_f_L}
\hat{f}_\mathcal{L}  \propto g_\mathcal{L}= f_{\mathcal{L}} \big|_{\hat{b},~ \phi \leq \text{max}(\phi)}
\end{equation}

with scaling parameter $\hat{b}=b\cdot \nu=\sigma_\phi/\sqrt{2} \cdot \nu $ and $\nu=1.32$ inducing same shape (slope and spread) as the numerical form in \eqref{eq:pdf_f_num} within $1\tfrac{1}{2}\%$ relative deviation to \eqref{eq:pdf_f_num}, see Fig. \ref{fig:circ_footprint}.

The 2D rotation symmetric distribution $f_{\mathrm{2D}}(r)$, has a polar sense variable $r\equiv\sqrt{\theta_{\mathrm{2D}}^2+\phi_{\mathrm{2D}}^2}$ ($\theta_{\mathrm{2D}}$ and $\phi_{\mathrm{2D}}$ interpreted as equivalent Cartesian coordinates), truncated at $r_0=\mathrm{max}(|\phi|)=\sigma_{\mathrm{rel}}\mathrm{max}(|\phi_{\mathrm{nom}}|)$.
However, there does not seem to be a simple tractable solution under the constraint $\int f_{\mathrm{2D}} d\theta_{\mathrm{2D}}=f_\mathcal{L}$.   Instead we use an approximation

\begin{equation}\label{eq:pdf_2D_L}
\hat{f}_{\mathrm{2D} \mathcal{L}}\propto g_{\mathrm{2D}\, \mathcal{L}}=\mathrm{exp}(-\sqrt{r}/b_{\mathrm{2D}})\big|_{ r \leq r_0}
\end{equation}

where $b_{\mathrm{2D}}=\hat{b}/(\beta_{\mathcal{L}}\sqrt{ \sigma_{\mathrm{rel}}})$ (using $\beta_{\mathcal{L}}=2.29$) and the exponent of $r$ (i.e. $\sqrt{r}$) are chosen for simplicity and good shoulder fit. After $\theta_{\mathrm{2D}}$ integration, we get a relative deviation to \eqref{eq:pdf_f_num}$  <5\%$  within the central ca. $3/4$ $\phi_{\mathrm{2D}}$-range, see  Fig. \ref{fig:circ_footprint}.

Analytical integration of \eqref{eq:pdf_2D_L} to facilitate normalization $\int  f=1$, seems difficult if at all tractable. Instead we choose a simple numerical normalization $\hat{f} \equiv g/\Sigma g$.

The grid points $\Omega_g=[(\theta_g,\phi_g)]$  inside the footprint, are found via a brute force 3D Pythagorean radius search as $[(\theta_g,\phi_g)] = \{(\theta,\phi)\in\Omega | ~r_{\mathrm{3D}} \leq r_{0~\mathrm{3D}}\}$. $r_{0~\mathrm{3D}}=\sqrt{2(1-\cos(r_0))}$  is the chord (3D Cartesian radius) of the circular footprint, while  $r_{\mathrm{3D}}=\sqrt{(x-x_c)^2+(y-y_c)^2+(z-z_c)^2}$ is the chord from circle center ($\theta_c,\phi_c$) to any point $(x,y,z)$ on the unit sphere $\Omega$. The Cartesian center coordinates are $x_c=\sin(\theta_c)\cos(\phi_c)$, $ y_c=\sin(\theta_c)\sin(\phi_c)$ and $ z_c=\cos(\theta_c)$. 

For larger footprints, the 3D spherical area element $d\Omega=\sin(\theta) d\theta d\phi$ compression  becomes noticeable.

However, we can facilitate the desired azimuth Laplacian approximation, under spherical $ \sin(\theta)d\theta$ integration with 3D arc $r_\Omega$, via a simple porting of \eqref{eq:pdf_2D_L} as
\begin{equation}\label{eq:pdf_3D_L}
\begin{cases}
r_\Omega=&\frac{180}{\pi} 2\arcsin(\frac{r_{\mathrm{3D}}}{2}) [^o]
\\
\hat{f}_{\mathrm{3D} \mathcal{L}}\propto g_{\mathrm{3D}\, \mathcal{L}} =&\mathrm{exp}\left (-\sqrt{r_{\Omega}}/b_{\mathrm{2D}} \right)\big|_{ r_{\mathrm{3D}} \leq r_{0\, \mathrm{3D}}}
\end{cases}
\end{equation}

This preserves good fit up to hemispherical\footnote{we cannot maintain circular footprints without $\theta$ vs. $\phi$ interpretation ambiguity, when crossing the hemisphere as we then get $\theta<0 \vee \theta>\pi$}  footprints  ($\sigma_{\mathrm{rel}}\lesssim 49$). The lower tails of the integrated azimuth response gets emphasized compared to \eqref{eq:pdf_2D_L} when approaching hemispherical coverage, see  Fig. \ref{fig:circ_footprint}. 

The grid point probability is weighted with its associated surface area.
The area of a unit sphere grid element is $\Delta A\approx\sin(\theta_g)\Delta\theta\Delta\phi$, with grid spacings $\Delta\theta$ and $\Delta\phi$.
For each pole ($\theta=0 \vee \pi$) we include the total cap area at the pole, as $ \sum \Delta A (\theta_g=0 \vee \pi) \equiv \frac{\gamma}{2}2\pi(1-\cos(\Delta\theta /2))$.  With tuning parameter $\gamma=1.337$  we achieve a good total area fit\footnote{Two orders of magnitude less relative approximation error (order $\epsilon_{rel}\sim 10^{-6}$ for grid spacings of \cite{ref:IEEE_access_SSZ_GFP, ref:IET_SSZ_AT_EF_GFP}) to the unit sphere area $A_\Omega=4\pi$, than when omitting top and bottom cap.}.

Finally, the footprint power density becomes a grid area weighted version of \eqref{eq:pdf_3D_L}
\begin{equation}
p_S(\Omega_g) \equiv \hat{f}_{\mathrm{3D}\, \mathcal{ L},w} =\frac{ g_{\mathrm{3D}\, \mathcal{L}} (\Omega_g)\Delta A (\Omega_g)}{\sum g_{\mathrm{3D}\, \mathcal{L}} (\Omega_g) \Delta A(\Omega_g) } 
\end{equation}

\section*{Acknowledgment}
We thank Kristian Bank for remeasuring, to a high density, the complex patterns on the mockup phone \cite{ref:IET_SSZ_AT_EF_GFP} at same bands used in \cite{ref:IEEE_access_SSZ_GFP} - and Gert Fr\o lund Pedersen for fruitful discussions on handset antenna behaviors.

\bibliographystyle{IEEEtran}

%\begin{IEEEbiography}{Patrick C.F. Eggers}
%Biography text here.
%\end{IEEEbiography}

%\begin{IEEEbiography}{Stanislav S. Zhekov}
%Biography text here.
%\end{IEEEbiography}

% insert where needed to balance the two columns on the last page with
% biographies
%\newpage

% You can push biographies down or up by placing
% a \vfill before or after them. The appropriate
% use of \vfill depends on what kind of text is
% on the last page and whether or not the columns
% are being equalized.

%\vfill

% Can be used to pull up biographies so that the bottom of the last one
% is flush with the other column.
%\enlargethispage{-5in}

\end{document}